\documentclass[11pt]{article}
\usepackage{latexsym}  
\usepackage{amssymb}
\usepackage{graphicx}
\usepackage{amsmath}

\begin{document}

\begin{center}

{\Large\bf Quark and Lepton Mass Matrix Model} \\[2mm]

{\Large\bf with Only Six Family-Independent Parameters}

\vspace{4mm}

{\bf Yoshio Koide$^a$ and Hiroyuki Nishiura$^b$}

${}^a$ {\it Department of Physics, Osaka University, 
Toyonaka, Osaka 560-0043, Japan} \\
{\it E-mail address: koide@kuno-g.phys.sci.osaka-u.ac.jp}

${}^b$ {\it Faculty of Information Science and Technology, 
Osaka Institute of Technology, 
Hirakata, Osaka 573-0196, Japan}\\
{\it E-mail address: hiroyuki.nishiura@oit.ac.jp}

\date{\today}
\end{center}

\vspace{3mm}

\begin{abstract}
We propose a unified mass matrix model for quarks and leptons, 
in which sixteen observables of mass ratios and mixings of 
the quarks and neutrinos are 
described by using no family number-dependent parameters except 
for the charged lepton masses 
and only six family number-independent free parameters.  
The model is constructed by extending the so-called ``Yukawaon" 
model to a seesaw type model.  
As a result, once the six parameters is fixed by the quark mixing 
and the mass ratios of quarks and neutrinos, no free parameters are 
left in the lepton mixing matrix.
The results are in excellent agreement with the neutrino mixing data.
We predict $\delta_{CP}^\ell =-68^\circ$ for the leptonic $CP$ 
violating phase and $\langle m\rangle\simeq 21$ meV for the 
effective Majorana neutrino mass. 
\end{abstract}

PCAC numbers:  
  11.30.Hv, 
  12.15.Ff, 
  14.60.Pq,  
  12.60.-i, 

\vspace{3mm}

\noindent{\it 1  \  Introduction}: 
It is a big concern in the flavor physics to investigate the origin of 
the observed hierarchical structures of  masses 
and mixings of quarks and leptons.
In the present paper, we attempt to describe the quark and neutrino  
mass matrices in terms of the charged 
lepton masses as only family number-dependent parameters 
with the help of the smallest number of possible family number-independent 
parameter.
We report in this paper that such an ambitious attempt has succeeded, and 
what is more surprising, the number of family 
number-independent free parameters of the model is only six
for sixteen observables.  
It should be noted that even in the latest Yukawaon model 
\cite{Yukawaon_PRD15} we needed ten family 
number-independent parameters. 
The success suggests that hierarchical structures in all the  
quark and lepton mass matrices are caused by one common origin. 
This result will bring new light to understand the origin of 
flavors.

In the so-called Yukawaon model 
\cite{Yukawaon_09}, the Yukawa coupling constants 
are considered to be effective coupling constants $Y_f^{eff}$ 
which are given by 
vacuum expectation values (VEVs) of scalars (``Yukawaons")  
$Y_f$ with $3 \times 3$ components for each flavor $f$:
$$
(Y_f^{eff})_i^{\ j} = ({y_f}/{\Lambda}) \langle Y_f\rangle_i^{\ j}  
\ \ \ \ (f=u, d, \nu, e),
\eqno(1)
$$
where $\Lambda$ is an energy scale of the effective theory. 
Although the Yukawaon model is a kind of flavon models \cite{flavon}, 
all the flavons in the Yukawaon model are expressed 
by $3 \times 3$ components. 
The Yukawaon model is based on the basic concepts that  
 (i) the fundamental flavor basis is a basis in which 
charged lepton mass matrix $M_e$ is diagonal and 
(ii) fundamental parameters in the quarks and leptons 
are $ (\sqrt{m_e},  \sqrt{m_\mu}, \sqrt{m_\tau})$
[not $(m_e,  m_\mu, m_\tau)$].  
These concepts have been motivated by a phenomenological success of 
the charged lepton mass relation  \cite{K-mass} 
$m_e + m_\mu + m_\tau = (2/3) (\sqrt{m_e} + \sqrt{m_\mu} +
\sqrt{m_\tau})^2$. 

\vspace{2mm}

{\it 2 \ VEV relations}: 
At first, we propose a model in the present paper, in which 
VEVs of the Yukawaons $Y_f$ (would-be Dirac mass matrices) 
take a universal seesaw form as given by
$$
\langle \hat{Y}_f \rangle_i^{\ j} = k_f \ 
 \langle {\Phi}_{0f} \rangle_i^{\ \alpha} 
\langle (S_f)^{-1} \rangle_{\alpha}^{\ \beta} 
\langle \bar{\Phi}_{0f}^T \rangle_{\beta}^{\ j} 
\ \ \ \ (f=u, d, \nu, e) ,
\eqno(2)
$$
where VEV s of the fields $\Phi_{0f}$, $S_f$, $\Phi_0$, $P_f $, 
and so on  are defined by 
$$
\langle \Phi_{0f}  \rangle_i^\alpha = ({1}/{\Lambda}) 
\langle \Phi_0 \rangle_{ik} 
\langle \bar{P}_f \rangle^{k\alpha},
\eqno(3)
$$
$$
\langle P_u  \rangle= v_P \, {\rm diag} (e^{i \phi_1},\ e^{i \phi_2},\, 
e^{i \phi_3}),
\ \ \ \langle  P_d \rangle = v_P {\bf 1} ,  \ \ \ 
\langle  P_\nu  \rangle= v_P {\bf 1}, \ \ \ 
\ \ \ \langle  P_e \rangle = v_P {\bf 1} ,
\eqno(4)
$$
$$
\langle \Phi_0 \rangle = v_0 \, {\rm diag}\, (z_1, z_2, z_3)
 \propto {\rm diag}\, 
(\sqrt{m_e}, \sqrt{m_\mu},  \sqrt{m_\tau}), 
\eqno(5)
$$
$$
\langle S_f \rangle =  v_{Sf} \left({\bf 1} + b_f e^{i\beta_f} 
X_3 \right) .
\eqno(6)
$$
Here $(m_{e1}, m_{e2}, m_{e3}) \equiv (m_e,m_\nu,m_\tau)$.  ${\bf 1}$ 
and $X_3$ are defined by
$$
{\bf 1} = \left( 
\begin{array}{ccc}
1 & 0 & 0 \\
0 & 1 & 0 \\
0 & 0 & 1 
\end{array} \right) , \ \ \ \ \ 
X_3 = \frac{1}{3} \left( 
\begin{array}{ccc}
1 & 1 & 1 \\
1 & 1 & 1 \\
1 & 1 & 1 
\end{array} \right) . 
\eqno(7)
$$
Here, we have assumed U(3)$\times$U(3)$'$ family-symmetries,
and indexes $i, j, \cdots$ and $\alpha, \beta, \cdots$ 
denote those of U(3) and U(3)$'$, respectively. 
The VEV form of $S_f$, Eq.(6), breaks the symmetry 
U(3)$'$ into a discrete symmetry S$_3$.  
The factor $S_f^{-1}$ in Eq.(2) comes from a seesaw scenario 
discussed in detail in Sec.3.
Especially, U(3)$'$ plays an essential role in considering 
a famly gauge boson model \cite{FGB}, in which masses of 
family gauge bosons $A_i^{\ j}$ are given by VEV of 
$(\Phi_0)_i^{\ \alpha}$, so that the model can avoid 
a severe constraint from the observed 
$K^0$-$\bar{K}^0$ mixing. 

In this model, a VEV form which explicitly breaks U(3) family 
symmetry is only the form $\langle \Phi_0\rangle$, Eq.(5). 
In this paper, we do not discuss the origin of the values 
$(z_1, z_2, z_3)$ given by Eq.(5), which is a basic assumption in the Yukawaon model.


A neutrino mass matrix is assumed, by adopting the conventional 
seesaw mechanism, as
$$
(M_\nu^{Majorana})_{ij} =\langle \hat{Y}_\nu \rangle_i^{\ k} 
\langle \bar{Y}_R^{-1} \rangle_{kl} 
\langle \hat{Y}_\nu^T \rangle^l_{\ j} .
\eqno(8) 
$$
Here, the following VEV structure of the $Y_R$ (the Majorana mass matrix
of the right-handed neutrinos $\nu_R$) is assumed, according to 
the previous Yukawaon model \cite{Yukawaon_PRD15}:
$$
\langle \bar{Y}_R \rangle^{ij} =k_R \frac{1}{\Lambda} \left[  
\left( \langle \bar{\Phi}_0 \rangle^{ik}
\langle \hat{Y}_u \rangle_k^{\ j}
 + \langle \hat{Y}_u^T \rangle^i_{\ k} \langle \bar{\Phi}_0 
\rangle^{kj} \right)
+  \frac{\xi_R}{\Lambda} \langle \hat{Y}_e^T \rangle^i_{\ k} 
\langle \bar{E} \rangle^{kl} 
\langle \hat{Y}_e \rangle_l^{\ j} \right] ,
 \eqno(9)
$$
where $\langle E\rangle = v_E {\bf 1}$.
Here, the last term ($\xi_R$ term) has been introduced in order to 
give a reasonable value for a neutrino mass squared difference ratio 
$R_\nu= (m_{\nu 2} ^2- m_{\nu 1}^2)/(m_{\nu 3}^2 -m_{\nu 2}^2)$. 
Exactly speaking, three terms  
$ \bar{E}^{ik} (\hat{Y}_e)_k^{\ l}  (\hat{Y}_e)_k^{\ j}$, 
$(\hat{Y}_e^T)^i_{\ k} \bar{E}^{kl} (\hat{Y}_e)_l^{\ j}$, and
$(\hat{Y}_e^T)^i_{\ k} (\hat{Y}_e^T)^k_{\ l} \bar{E}^{lj}$ 
are possible as the $\xi_R$ term. 
However, since we have considered 
$\langle E\rangle = v_E {\bf 1}$, 
we have denoted only one term of the three terms in Eq.(9) 
for convenience.  

Since we deal with mass ratios and mixings only, the common 
coefficients $k_f$, $v_{Sf}$, and so on does not 
affect the numerical results, so that hereafter we omit such
coefficients even it those have dimensions. 

Let us state some remarks on our new Yukawaon model in order:

(a) The VEV form (5) is a fundamental postulation in the 
Yukawaon model. 
We assume that the VEV form of $\Phi_0$ is diagonal in the 
flavor basis in which  $\langle S_f\rangle$ take a form
''unit matrix plus democratic matrix".
We do not ask the origin of the values of $z_i$ for the moment.   

(b) The structures of the quark and  Dirac neutrino mass matrices 
$\langle\hat{Y}_f\rangle$ are essentially determined by 
the parameters  $b_f e^{i\beta_f} $.
Since we take a superpotential
$$
W_S= \mu_{1S} {\rm Tr}[S_e S_\nu] +  \mu_{2S} {\rm Tr}[S_e]  
{\rm Tr}[S_\nu] ,
\eqno(10)
$$ 
which leads to $\langle S_e \rangle =v_S {\bf 1}$ 
and $\langle S_\nu \rangle =v_S {\bf 1}$,
we obtain $b_e =b_\nu=0$ in the lepton sector. So that 
$\langle\hat{Y}_e\rangle$ and  $\langle\hat{Y}_\nu\rangle$ 
are given by a common form 
$\langle \Phi_0 \rangle  \langle \Phi_0 \rangle$.  
(However, this does not mean that $\hat{Y}_e$ and  
$\hat{Y}_\nu$ are a comon one flavon.) 
Of course, here, we assume $R$ charges, $R(S_\nu) + R(S_e)=2$. 

(c) VEV relations are derived from 
the supersymmetric vacuum conditions.
The possible combinations among those flavons are 
selected by $R$ charges in the SUSY scenario. 
See, for instance, Ref.\cite{Yukawaon_PRD14}. 
For example, the forms (4) are derived superpotential terms
$$
\begin{array}{l}
W_{Pq} = ({1}/{\Lambda}) \left( \lambda_{1Pq} 
{\rm Tr} [P_u \bar{P}_u P_d \bar{P}_d]
+  \lambda_{2Pq} {\rm Tr} [P_u \bar{P}_u ]
{\rm Tr} [P_d \bar{P}_d]  \right) , \\
W_{P\ell} = ({1}/{\Lambda}) \left( \lambda_{1P\ell} 
{\rm Tr} [P_\nu \bar{P}_\nu P_e \bar{P}_e]
+   \lambda_{2P\ell} {\rm Tr} [P_\nu \bar{P}_\nu ] 
{\rm Tr} [P_e \bar{P}_e] \right), \\
\end{array}
\eqno(11)
$$
which lead to VEV relations 
$\langle P_f\rangle \langle\bar{P}_f\rangle =v_P^2 {\bf 1}$ 
($f=u, d, \nu, e$).   
We regard the VEV form (4) 
as one of special solutions in 
the general relation 
$\langle P_f\rangle \langle \bar{P}_f\rangle =v_P^2 {\bf 1}$.   
(Here, we have taken  $R$ charge relations
$R(P_u)+R(P_d)=1$ and $R(P_\nu)+R(P_e)=1$.)

(d) The parameters $\phi_i$ ($i=1,2,3$) in Eq.(4) look like typical family 
number-dependent parameters. 
However, in the previous Yukawaon model \cite{Yukawaon_PRD15}, 
we have proposed a mechanism that the parameters $\phi_i$ can 
always be expressed in terms of the charged lepton mass 
parameters $m_{ei}$ with the help of two family number-independent 
parameters. 
The reason is as follows: when we put 
$(\phi_1, \phi_2, \phi_3)=(\phi_0 +\tilde{\phi}_1,
\phi_0 +\tilde{\phi}_2, \phi_0)$, the phase values 
$(\tilde{\phi}_1, \tilde{\phi}_2)$ are observables in 
fitting of the 
Cabibbo-Kobayashi-Maskawa (CKM) \cite{CKM} mixing 
parameters, but $\phi_0$ is not observable.
Therefore, we can always relate the values 
$(\phi_1, \phi_2, \phi_3)$ to the values  
 $(m_e, m_\mu, m_\tau)$ by adjusting $\phi_0$.
Therefore, for convenience, we count the parameters 
 $(\tilde{\phi}_1, \tilde{\phi}_2)$ as family 
number-independent parameters.
 (For the details, see Ref.\cite{Yukawaon_PRD15}.)

(e) In the previous model \cite{Yukawaon_PRD15}, we have discussed
VEV form $P_f \Phi_0 S_f \Phi_0 P_f$ with 
$S_f ={\bf 1}+ a_f e^{i\alpha_f} X_3$ (not $\Phi_0 S_f^{-1} \Phi_0$), in which 
we have taken the cases that $(\alpha_u =0, P_u\neq {\bf 1})$ and 
$(\alpha_d \neq 0, P_d ={\bf 1})$. 
The result comes from a rule that all VEV matrices of flavons satisfy 
$\langle \bar{A} \rangle = \langle A \rangle$  except for 
$\langle \bar{P} \rangle = \langle P \rangle^*$ with $\phi_i\neq 0$. 
However, in the present model, the corresponding VEV form is 
$\bar{P}_f \Phi_0 S_f \bar{\Phi}_0 P_f$, not 
$\bar{P}_f \Phi_0 S_f {\Phi}_0 {P}_f$, 
so that we cannot obtain a similar result as in 
Ref.\cite{Yukawaon_PRD15}.
Therefore, in the present paper, the relations 
$(\beta_u =0, P_u\neq {\bf 1})$ and $(\beta_d \neq 0, P_d ={\bf 1})$ 
are only assumptions. 
This is a task in future.

Finally, we summarize the parameters in the present model.  
We have only six free parameters
 $b_u$, $b_d$, $\beta_d$, 
$(\tilde{\phi}_1, \tilde{\phi}_2)$, $\xi_R$ for sixteen observable 
quantities (four quark mass ratios, two neutrino mass 
ratios, four CKM mixing parameters and four plus two 
 Pontecorvo-Maki-Nakagawa-Sakata (PMNS) \cite{PMNS}
mixing parameters). 
(Hereafter, for convenience, we denote 
$(\tilde{\phi}_1, \tilde{\phi}_2)$ as $({\phi}_1, {\phi}_2)$ simply.)
Note that the number of parameters is surprisingly small.
The parameter $b_u$ gives the up-quark mass ratios $m_u/m_c$
and $m_c/m_t$, so that we have one prediction.
The parameters $b_d$ and $\beta_d$ are fixed by the observed 
down-quark mass rations. 
Therefore, four parameters in the CKM mixing matrix are 
described only by two parameters $(\phi_1, \phi_2)$. 
The parameter $\xi_R$ adjusts the neutrino mass ratio $R_\nu$.
If the value $\xi_R$ is fixed, four plus two lepton mixing parameters 
are predicted with no free parameters.  
However, since the neutrino mass ratio has not been so precisely  
measured at present, we will give predictions of  two neutrino 
mass ratios and four PMNS mixing parameters (and with two Majorana 
phase) by adjusting one parameter $\xi_R$ as a practical matter. 
(See Fig.2.)


\vspace{2mm}

\noindent{\it 3 \ Seesaw-type mass matrix model}:
%
Let us discuss the origin of the VEV structure in Eq.~(2). 
The mass matrix model given in Eq.~(2)  is a sort of seesaw type 
mass matrix model for all the flavor. 
The form (2) has been suggested by a block diagonalization of 
a $6\times 6$ mass matrix term in a universal seesaw model given by 
$$
(\bar{f}_L^i \ \ \bar{F}_L^\alpha ) 
\left(
\begin{array}{cc}
(0)_i^{\ j}  &  \langle \Phi_{0f} \rangle_i^{\ \beta}  \\
\langle \bar{\Phi}_{0f}^T\rangle_\alpha^{\ j} & 
-\langle S_f\rangle_\alpha^{\ \beta} 
\end{array} \right) 
 \left(
\begin{array}{c}
f_{Rj} \\
F_{R\beta}
\end{array} 
\right) .
\eqno(12)
$$
Here $f_{L(R)}$ and $F_{L(R)}$ are, respectively, 
left (right) handed light and heavy  fermions fields. 
Exactly speaking, we have to read $\bar{f}_L$ in Eq.(13) as 
$\bar{f}_L H_{u/d}/\Lambda$. 
However, for convenience, we have denoted
those as $\bar{f}_L$ simply.
The mass matrix model with the heavy fermion mass matrix 
$M_F = - S_f$ with the form (6) is known as the so-called 
``democratic seesaw" model\cite{YK-HF_ZPC96}. 
The authors in Ref.\cite{YK-HF_ZPC96} have considered that 
the observed top quark mass enhancement originates in a condition 
${\rm det}M_F=0$ in the seesaw mass matrix (12).
In the seesaw approximation form (2), the condition 
${\rm det}M_F \rightarrow 0$ gives $m_3 \rightarrow \infty$ 
for one of the mass eigenvalues $(m_1,m_2,m_3)$. 
However, they found that the exact diagonalization of $6\times 6$ 
mass matrix (12) gives $m_3 \sim \Lambda_{weak}$, 
for one of the mass eigenvalues $(m_1,m_2,m_3, m_4, m_5, m_6)$,  
where $\Lambda_{weak}$ is a breaking scale of the electroweak symmetry.
The reason is quite simple: 
The matrix (6) with $f=u$ takes $S_u = {\rm diag}(1,\ 1, \ 0)$ 
in the limit of $b_u \rightarrow -1$ 
(i.e. ${\rm det}S_u \rightarrow 0$), 
so that the seesaw suppression affects only the first and second 
generations of up-quarks, so that the third generation quark $t$ 
takes mass of the order of $\Lambda_{weak}$ without the seesaw 
suppression. 
Furthermore, the model can give $m_u \sim  m_d \sim m_e$
insensitively to the values of the parameters $b_u$ and $b_d$ 
as seen in Sec.4 later.
In spite of such successful description of quark masses and mixing, 
the authors in Ref.\cite{YK-HF_ZPC96} failed to give reasonable 
neutrino masses and mixing. 
On the other hand, the Yukawaon model have succeeded in giving 
 not only reasonable quark mass ratios and mixing but also 
neutrino masses and mixing. 
However, the Yukawaon model needs a lot of parameters.
In the present paper, we have applied this seesaw model 
to our Yukawaon model.  

However, note that the $3\times 3$ mass matrix between $\bar{f}_L$ 
and $f_R$ is absent in the form (12).
If we assume the seesaw mechanism plus the Yukawaon model,
$$
(\bar{f}_L^i \ \ \bar{F}_L^\alpha ) 
\left(
\begin{array}{cc}
(\hat{Y}_f)_i^{\ j}  &  (\Phi_{0f})_i^{\ \beta}  \\
(\bar{\Phi}_{0f}^T)_\alpha^{\ j} & -(S_f)_\alpha^{\ \beta} 
\end{array} \right) 
 \left(
\begin{array}{c}
f_{Rj} \\
F_{R\beta}
\end{array} 
\right) ,
\eqno(13)
$$
then, we obtain a $3\times 3$ mass matrix between  
$\bar{f}'_L$ and $f'_R$, 
$$
M_f \simeq \hat{Y}_f + \Phi_{0f} S_f^{-1} \bar{\Phi}_{0f} ,
\eqno(14)
$$
after the block diagonalization. 
(Here and hereafter, for convenience, we sometimes omit the notations 
``$\langle$" and ``$\rangle$" which denote VEV matrices.)  
However, note that, in Eq.(14), 
the first term $\hat{Y}_f$ is independent of the second
term $ \Phi_{0f} S_f^{-1} \bar{\Phi}_{0f}$. 
In order to obtain the relation (2), 
we put the following two assumptions:

\noindent
[Assumption 1]  The VEV value $\hat{Y}_f$ and the VEV value
$M_F= -S_f$ take the same scale transformation (we denote 
the scale transformation as a parameter $\zeta_f$):
$$
 M_f =  \zeta_f \hat{Y}_f + ({1}/{\zeta_f} )\Phi_{0f} S_f^{-1}
 \bar{\Phi}_{0f} .
\eqno(15)
$$
[Assumption 2] The VEV value $\hat{Y}_f$ is taken so that  
$M_f$ takes a locally minimum value under the $\zeta_f$ transformation:
$$
\frac{\partial M_f}{\partial \zeta_f} = \hat{Y}_f -\frac{1}{\zeta_f^2}
 \Phi_{0f} S_f^{-1}  \bar{\Phi}_{0f} =0 .
\eqno(16)
$$  
Then, we obtain
$$
\hat{Y}_f = ({1}/{\zeta_f^2}) \Phi_{0f} S_f^{-1} \bar{\Phi}_{0f} ,
\ \ {\rm i.e.} \ \
M_f = ({2}/{\zeta_f}) \Phi_{0f} S_f^{-1} \bar{\Phi}_{0f} 
= 2 \zeta \hat{Y}_f.
\eqno(17)
$$

In Ref.\cite{YK-HF_ZPC96}, the up-quark masses $m_{ui}$ 
have been estimated by diagonalizing $6\times 6$ mass matrix (12) 
with the input value $b_u = -1$ with $\beta_u=0$.
However, in this paper, for convenience, we use the approximate  
expression (17), although the seesaw approximation (17) is not 
valid for $b_u=-1$.
Therefore, instead of $b_u= -1$, the parameter value $b_u$ is fixed by 
the observed value of the up-quark mass ratio $m_c/m_t$. 
(We will obtain $b_u= -1.011$ as seen in Sec.4.)
We use the relation (17) for Dirac masses of 
all quarks and leptons. 
(Hereafter, we put simply $\zeta_f=1$.)

\vspace{2mm}

\noindent{\it 4 \  Numerical predictions}: 
%
We summarize our mass matrices for the numerical analysis as follows:
$$
 \hat{Y}_u = \Phi_0 \bar{P}_u ({\bf 1} + b_u X_3)^{-1}
 P_u \bar{\Phi}_0 , \quad \hat{Y}_d = \Phi_0  ({\bf 1} + b_d e^{i\beta_d} X_3)^{-1}
 \bar{\Phi}_0 ,  \quad \hat{Y}_e = \Phi_0 \bar{\Phi}_0 , 
\eqno(18)
 $$
 $$
 M_\nu^{Majorana}=\hat{Y}_\nu \bar{Y}_R^{-1} \hat{Y}_\nu^T, 
\quad \bar{Y}_R=\bar{\Phi}_0 \hat{Y}_u +\hat{Y}_u^T
\bar{\Phi}_0+\xi_R \hat{Y}_e^T \hat{Y}_e, \quad
\hat{Y}_\nu = \Phi_0 \bar{\Phi}_0.
\eqno(19)
 $$
For convenience of numerical fitting, we re-define all VEV 
matrices of flavons as dimensionless matrices, i.e. 
$\bar{P}_u= {\rm diag}(e^{i \phi_1}, e^{i  \phi_2}, 1)$,  
$\Phi_0 = {\rm diag}(z_1, z_2, z_3)$, and so on. 
For the input values  $m_{ei}$, 
we use the values at $\mu=m_Z$.  
For the parameter $\xi_R$ defined in $\bar{Y}_R$, 
we redefine ${\xi_R}/{\Lambda}$ as  $\xi_R$.

The parameter value of $b_u$ can be determined from the observed
 up-quark mass ratio $m_c/m_t$ at $\mu= m_Z$.
we determine $b_u =-1.011$, which leads to the 
up-quark mass ratios   
$$
r^u_{12} \equiv \sqrt{{m_u}/{m_c} } = 0.061, \ \ \ \ 
r^u_{23} \equiv \sqrt{{m_u}/{m_c} } = 0.060 ,
\eqno(20)
$$
which are good in agreement with the observed up-quark mass ratios 
at $\mu= m_Z$ \cite{q-mass},  
$r^u_{12} = 0.045^{+0.013}_{-0.010}$ and
$r^u_{23}  =0.060 \pm 0.005$. 
Here, we have used the value of $r^u_{23}$ as an input value 
in determining $b_u$
because the light quark masses have large errors.
Although the predicted value $r^u_{12}$ in Eq.(20) is  somewhat large 
compared with the observed value,  we consider that 
this discrepancy is acceptable, since the purpose of the present paper 
is to give an overview of quark and lepton masses and mixings
with free parameters  as few as possible.

For down-quark mass ratios, we take parameter values 
$b_d = -3.3522$ and $\beta_d = 17.7^\circ$ which gives
down-quark mass ratios
$$
r^d_{12} \equiv {m_d}/{m_s} = 0.049, \ \ \ \ 
r^d_{23} \equiv {m_s}/{m_b} = 0.027 .
\eqno(21)
$$
The observed down-quark mass ratios at $\mu = m_Z$ are a little 
controversial:  
$r^d_{12} =  0.053^{+0.005}_{-0.003}$ and $r^d_{23}= 0.019 \pm 0.006$ 
(Xing, {\it et al}. \cite{q-mass}) and also 
$r^d_{12}= 0.050 \pm  0.010$ 
and $r^d_{23} = 0.031 \pm 0.004$ ({Fusaoka-Koide  \cite{q-mass2}). 
Our model cannot give the observed values by Xing {\t el al}. 
Our best fit parameter values are near to the values given 
in Ref.\cite{q-mass2}.
The fitting of $b_d$ and $\beta_d$ have been done with keeping 
in mind that it also leads to consistent CKM mixings, since 
the CKM mixings also depend on $b_d$ and $\beta_d$.

The explicit mass eigenvalues are as follows:
$(m_u, m_c, m_t) = (0.000398, 0.1064, 29.74) m_0$ and 
$(m_d, m_s, m_b)= (0.000725, 0.01467, 0.5365) m_0$, 
where $m_0 =(v_0 v_P/\Lambda)^2/v_S$ and $v_{Su} =v_{Sd}  
\equiv v_S$,  
so that we can obtain reasonable $m_d/m_u$ ratio,  
${m_d}/{m_u} = 1.8 $,  
which well agrees with the observed ratio \cite{q-mass2}
${m_d}/{m_u}  = 2.01^{+0.47}_{-0.46}$.
Though we obtain $(m_e. m_\mu, m_\tau) = 
m_0 (0.000263, 0.0555, 0.9442)$, the ratio 
$m_e/m_u \sim 0.66$ is not agree with the observe 
value $(m_e/m_u)^{obs} \sim 0.38$.
We think that the common coefficient $m_0$ should
be distinguished between $(m_0)_{quark}$ and 
$(m_0)_{lepton}$, and the difference is originated 
in the difference of $v_{Sq}\equiv v_{S}(quark)$ and 
$v_{S\ell} \equiv v_S(lepton)$ in Eq.(6).
It is interesting that this discrepancy suggests 
$v_{Sq}\simeq v_{S\ell}/\sqrt{3}$.

Next, let us try to fitting CKM mixing parameters.
Since the parameters  $b_u$, $b_d$ and $\beta_d$
have been fixed by the observed quark mass rations, 
the CKM mixing matrix elements  $|V_{us}|$, 
$|V_{cb}|$, $|V_{ub}|$, and  $|V_{td}|$ are functions of 
the remaining two free  parameters 
$\phi_1$ and $\phi_2$.
In Fig.~1, we draw contour curves of the CKM mixing matrix elements 
in the ($\phi_1$, $\phi_2$) 
parameter plane which are obtained from the observed constraints of 
the CKM mixing matrix elements, 
with taking  $b_u=-1.011$, and  $b_d=-3.3522$, $\beta_d=17.7^\circ$.  
%
As shown in Fig.~1, all the experimental constraints on 
CKM parameters are satisfied by 
fine tuning of the parameters $\phi_1$ and $\phi_2$ as 
$(\phi_1, \phi_2)=(-176.05^\circ, -167.91^\circ )$, 
which predicts 
$$
|V_{us}|= 0.2257 , \ \ \  |V_{cb}|= 0.03996 ,  \ \ \ |V_{ub}|= 0.003701 , 
\ \ \ |V_{td}|= 0.009173    , \ \ \  \delta_{\rm CP}^q=80.99   .
\eqno(22)
$$

\vspace{2mm}

\begin{figure}[ht]
\begin{picture}(200,200)(0,0)

  \includegraphics[height=.33\textheight]{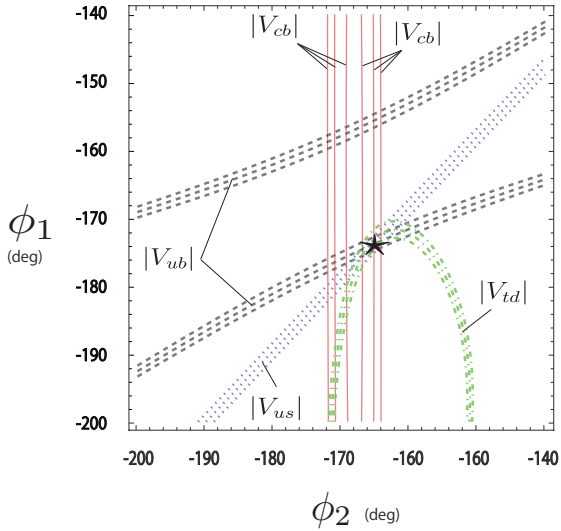}

\end{picture}  
  \caption{Contour curves  in the ($\phi_1$, $\phi_2$) 
parameter plane of the observed CKM mixing matrix elements of  $|V_{us}|$, 
$|V_{cb}|$, $|V_{ub}|$, and  $|V_{td}|$.
We draw the three contour curves, which corresponds to the center, 
upper, and lower values of the observed constraints for each  
the CKM mixing matrix elements,  
with taking  $b_u=-1.011$, and  $b_d=-3.3522$, $\beta_d=17.7^\circ$.
We find that the parameter set around ($\phi_1$, $\phi_2$) 
$=(-176.05^\circ, -167.91^\circ )$ 
indicated by a star ($\star$) is 
consistent with all the observed values.
} 
 \label{fig1}
\end{figure}

\begin{figure}[ht]
\begin{picture}(200,200)(0,0)

  \includegraphics[height=.33\textheight]{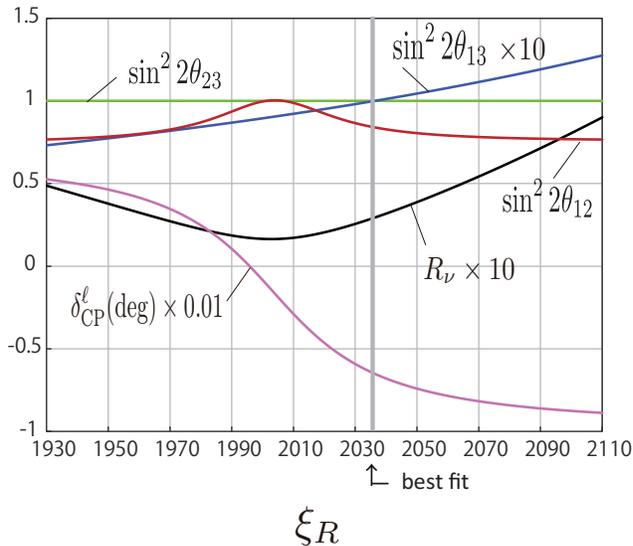}

\end{picture}  
  \caption{$\xi_R$ dependence of the lepton mixing parameters 
$\sin^2 2\theta_{12}$, 
$\sin^2 2\theta_{23}$, $\sin^2 2\theta_{13}$, and the neutrino 
mass squared difference ratio $R_\nu$. 
We draw curves of those as functions of $\xi_R$  
for the case of $b_u=-1.011$ and 
$(\phi_1, \phi_2)=(-176.05^\circ, -167.91^\circ )$.
} 
 \label{fig2}
\end{figure}

Now let us present the result for the neutrino sector.
Substantial differences between the present and previous papers appear 
in the parameter fitting of the PMNS lepton mixing. 
We have already fixed the four parameters  $b_u$, $b_d$,  
$\beta_d$, $\phi_1$, and 
$\phi_2$ from the quark mass ratios and CKM mixing.  
Therefore, 
the PMNS mixing parameters $\sin^2 2\theta_{12}$, $\sin^2 2\theta_{23}$, 
$\sin^2 2\theta_{13}$, 
$CP$ violating Dirac phase parameter $\delta_{\rm CP}^\ell$,  and 
the neutrino mass squared difference ratio 
$R_{\nu} \equiv {\Delta m_{21}^2}/{\Delta m_{32}^2}$  are 
turned out to be functions 
of the remaining only one  parameter $\xi_R$. 
In Fig.~2, we draw the curves as  functions of  $\xi_R$
with taking  $b_u=-1.011$, and  $b_d=-3.3522$, $\beta_d=17.7^\circ$, and 
$(\phi_1, \phi_2)=(-176.05^\circ, -167.91^\circ )$. 
As seen in Fig.2, we find that the predicted value of 
the $\sin^2 2\theta_{23}$ 
is almost constant as $\sin^2 2\theta_{23} \simeq 1$ and 
the value   $\sin^2 2\theta_{12}$ is not so sensitive to the
parameter $\xi_R$  as 
$\sin^2 2\theta_{12}= 0.8 -1.0$.
By using the observed value of $R_\nu$ as input, we determine the value  
$\xi_R = 2039.6$ in the unit of $v_e^2 v_E/v_o v_u \Lambda$, 
which gives the predictions
$$
\sin^2 2\theta_{12} = 0.8254, \ \ 
\sin^2 2\theta_{23} =0,9967, \ \ \sin^2 2\theta_{13} = 0.1007, \ \ 
 \delta_{\rm CP}^\ell =-68.1^\circ,  \ \  R_{\nu} = 0.03118.
\eqno(23)
$$
These predictions are in good agreement with
the observed values \cite{PDG14} given in Table 1

Now, we predict neutrino masses with a normal hierarchy, 
which are consistent with the observed oscillation data, as 
$$
m_{\nu 1} \simeq 0.038\ {\rm eV}, \ \ m_{\nu 2} \simeq 0.039 \ {\rm eV}, 
\ \ m_{\nu 3} \simeq 0.063 \ {\rm eV}  ,
\eqno(24)
$$
by using the input value \cite{PDG14}
$\Delta m^2_{32}\simeq 0.00244$ eV$^2$. 
We also predict the effective Majorana neutrino mass \cite{Doi1981} 
$\langle m \rangle$ 
in the neutrinoless double beta decay as
$$
\langle m \rangle =\left|m_{\nu 1} (U_{e1})^2 +m_{\nu 2} 
(U_{e2})^2 +m_{\nu 3} (U_{e3})^2\right| 
\simeq21 \times 10^{-3} \mbox{ eV}.
\eqno(25)
$$

The predictions of our model are listed in Table~1. 
The process for fitting parameters is summarized  in Table.~2.

\vspace{2mm}
\begin{table}
\caption{Predicted values vs. observed values. 
} 

\vspace*{2mm}
\hspace*{-6mm}
\begin{tabular}{|c|ccccccccc|} \hline
  & $|V_{us}|$ & $|V_{cb}|$ & $|V_{ub}|$ & $|V_{td}|$ & 
$\delta^q_{CP}$ &  $r^u_{12}$ & $r^u_{23}$ & $r^d_{12}$ & $r^d_{23}$ 
 \\ \hline 
Pred &$0.2257$ & $0.03996$ & $0.00370$ & $0.00917$ & $81.0^\circ$ & 
$0.061$ & $0.060$ & $0.049$ & $0.027$ 
 \\
Obs & $0.22536$ & $0.0414$ &  $0.00355$  & $0.00886$  & $69.4^\circ$ &
$0.045$ & $0.060$ & $0.053$  & $0.019$  
  \\ 
    &  $ \pm 0.00061$ &  $ \pm 0.0012$& $ \pm 0.00015$ & 
 ${}^{+0.00033}_{-0.00032}$ & $\pm 3.4^\circ$ &
${}^{+0.013}_{-0.010}$ & $ \pm 0.005$ & $^{+0.005}_{-0.003}$ &
${}^{+0.006}_{-0.006}$ 
 \\ \hline
   & $\sin^2 2\theta_{12}$ & $\sin^2 2\theta_{23}$ & $\sin^2 2\theta_{13}$ & 
 $R_{\nu}\ [10^{-2}]$ &  
$\delta^\ell_{CP}$ & $m_{\nu 1}\ [{\rm eV}]$ & $m_{\nu 2}\ [{\rm eV}]$ & 
$m_{\nu 3}\ [{\rm eV}]$ & $\langle m \rangle \ [{\rm eV}]$ \\ \hline
 Pred & $0.8254$ & $0.9967$ & $0.1007$ &  $3.118$ & $-68.1^\circ$ &
 $0.038$ & $0.039$ & $0.063$ & $0.021$ \\
Obs & $0.846$   & $ 0.999$ & $0.093$ &   $3.09 $    & -
  &  -  &  -  &  -  &  $<\mathrm{O}(10^{-1})$   \\ 
    & $ \pm 0.021$ & $^{+0.001}_{-0.018}$   & $\pm0.008$ & $ \pm 0.15 $  &  &
   &    &    &    \\ \hline 
\end{tabular}
\end{table}

\begin{table}
\caption{Process for fitting parameters. 
}
\vspace{2mm}
\begin{center}
\begin{tabular}{|c|cc|cc|c|} \hline
Step & Inputs & $N_{input}$ &  Parameters & $N_{parameter}$ &
 Predictions  \\ \hline
1st  
     & $m_c/m_t$ & 1 & $b_u$ & 1 & $m_u/m_c$ \\
     &  $m_d/m_s$, $m_s/m_b$ & 2 & $a_d$, $\beta_d$ & 2 & $m_d/m_u$  \\
2rd  & $|V_{us}|$, $|V_{cb}|$  & 2 &   $(\phi_1, \phi_2)$ & 2 & 
$|V_{ub}|$,  $|V_{td}|$, $\delta_{CP}^q$  \\
3rd  &  $R_\nu$  & 1 
& $\xi_R$  & 1 & $\sin^2 2\theta_{12}$, $\sin^2 2\theta_{23}$,
$\sin^2 2\theta_{13}$, $\delta_{CP}^\ell$  \\
    &    &   &   &  & 2 Majorana phases, $\frac{m_{\nu 1}}{m_{\nu 2}}$,
 $\frac{m_{\nu 2}}{m_{\nu 3}}$   \\ 

option &  $\Delta m^2_{32}$ &   & $m_{\nu 3}$ &  & 
$(m_{\nu 1}, m_{\nu 2},  m_{\nu 3})$, $\langle m \rangle$  \\
\hline
$\sum N_{total}$ & & 6 &  & 6 &   \\ \hline 
\end{tabular}
\end{center}
\end{table}

\vspace{2mm}

\noindent{\it 5 \  Concluding remarks}:  %
We have proposed a model which combines the Yukawaon 
model \cite{Yukawaon_PRD15} with 
the democratic seesaw scenario \cite{YK-HF_ZPC96}, 
and have  demonstrated that 
the observed masses and mixings of quarks and 
neutrinos can be described only by the observed charged 
lepton mass values   
and six family number-independent parameters. 
The model provides an interesting picture that the observed hierarchical 
structures of the masses and the mixings of quarks and neutrinos
are brought by a common origin which comes from the charged lepton 
mass spectrum $(m_e, m_\mu, m_\tau)$. 
The model also predicts $\delta_{CP}^\ell = -68^\circ$ 
for the leptonic $CP$ violating Dirac phase, which will be checked 
by neutrino oscillation experiments in the near future. 
The prediction $\langle m\rangle \simeq 21$ meV is also  
within the reach of neutrinoless double beta decay experiments 
in the near future. 

For a full table of our flavons together with their $R$ charges 
and full expressions of  all the superpotential terms, 
we will give them elsewhere.

\vspace{5mm}
%

\end{document}